\documentclass[prl,showpacs,twocolumn,preprintnumbers,amsmath,groupedaddress,amssymb,superscriptaddress,floatfix]{revtex4}
\usepackage[dvips]{graphicx}
\usepackage[latin1]{inputenc}
\usepackage{xcolor,bm}
\usepackage{nicefrac}

\begin{document}
\draft

\hyphenation{a-long}

\title{Monopole-limited nucleation of magnetism in Eu$_{\bf{2}}$Ir$_{\bf{2}}$O$_{\bf{7}}$}

\author{Giacomo~Prando}\email[E-mail: ]{giacomo.prando@unipv.it}\affiliation{Department of Physics, University of Pavia, I-27100 Pavia, Italy}
\author{Prachi Telang}\affiliation{Department of Physics, Indian Institute of Science Education and Research, Pune 411 008, India}
\author{Stephen D. Wilson}\affiliation{Department of Materials, University of California, Santa Barbara, California 93106, USA}
\author{Michael J. Graf}\affiliation{Department of Physics, Boston College, Chestnut Hill, Massachusetts 02467, USA}
\author{Surjeet Singh}\affiliation{Department of Physics, Indian Institute of Science Education and Research, Pune 411 008, India}\affiliation{Center for Energy Sciences, Indian Institute of Science Education and Research, Pune 411 008, India}

\widetext

\begin{abstract}
	We present an in-depth analysis of muon-spin spectroscopy measurements of Eu$_{2}$Ir$_{2}$O$_{7}$ under the effect of the Eu$_{1-x}$Bi$_{x}$ isovalent and diamagnetic substitution as well as of external pressure. Our results evidence an anomalously slow increase of the magnetic volume fraction upon decreasing temperature only for stoichiometric Eu$_{2}$Ir$_{2}$O$_{7}$, pointing towards highly unconventional properties of the magnetic phase developing therein. We argue that magnetism in Eu$_{2}$Ir$_{2}$O$_{7}$ develops based on the nucleation of magnetic droplets at $T_{N}$, whose successive growth is limited by the need of a continuous generation of magnetic hedgehog monopoles.
\end{abstract}

\date{\today}

\maketitle

\narrowtext

\section{Introduction}

Pyrochlore iridium oxides -- with characteristic chemical formula $R_{2}$Ir$_{2}$O$_{7}$, $R$ being a rare-earth ion -- host a wide variety of exotic electronic phases arising from the combined effect of the spin-orbit interaction, Coulombic correlations and the peculiar geometrical properties of the lattice \cite{Gar10,Shi19} making them candidate materials for the realization of electronic states with non-trivial topological properties \cite{Pes10,Wan11,Moo13,Wit14,Sus15,Gos17,Ber18}. Particular interest has been devoted to the metal-insulator transition developing for $R$ = Eu, Sm and Nd, where a low-temperature magnetic insulating phase evolves into a non-magnetic metallic state above a critical temperature $T_{N}$ \cite{Tai01,Yan01,Mat07,Mat11,Zha11,Dis12,Ish12,Tom12,Dis13,Guo13,Nak16,Asi17,Zha17}. Interestingly, $T_{N}$ shows a marked decrease upon increasing the average ionic size at the $R$ site towards the limit of Pr$_{2}$Ir$_{2}$O$_{7}$, a metallic spin liquid which remains non-magnetic down to the lowest accessible temperatures \cite{Nak06,Mac07,Tok14}. Theoretical proposals suggest that the complete suppression of $T_{N}$ takes place at a quantum critical point whose properties are closely tied with those of the magnetic state of the iridium sublattice \cite{Sav14}, where the magnetic moments are forced along the local $\langle 1 1 1 \rangle$ directions pointing all inwards or outwards the tetrahedron defining their crystallographic sites, realizing the so-called all-in--all-out order \cite{Ari13,Sag13,Dis14,Don16}.

Strategies for a gradual approach to the quantum critical point include progressive chemical substitutions at the $R$ site as well as the application of external pressure \cite{Sak11,Taf12,Ued15,Pra16}. In this respect, a promising family of materials is $\left(\textrm{Eu}_{1-x}\textrm{Bi}_{x}\right)_{2}$Ir$_{2}$O$_{7}$ thanks to the isovalent character of the chemical substitution and to the diamagnetic nature of both Eu$^{3+}$ and Bi$^{3+}$ ions. The latter property is particularly relevant as it makes the intrinsic magnetic state of the iridium sublattice accessible without complications due to the $f-d$ exchange. Recently, it was reported that $10$ \% of bismuth in the system is enough to suppress magnetism and to induce a linear-in-temperature dependence for resistivity - a possible signature of quantum criticality in the system \cite{Tel19}.

In this work, we report on a study of Eu$_{2}$Ir$_{2}$O$_{7}$ as performed by means of muon-spin spectroscopy ($\mu^{+}$SR). We follow the evolution of the magnetic ground state towards its full suppression at different values $x$ for the Eu$_{1-x}$Bi$_{x}$ chemical substitution. Focusing on the dependence of the magnetic volume fraction $V_{m}$ of the sample on temperature, we highlight a highly unusual behaviour specific of the stoichiometric unsubstituted composition. We argue that the observed behaviour should be ascribed to the all-in--all-out order and to the peculiar topological properties of magnetically-ordered tetrahedra.

\section{Experimental details}

We performed $\mu^{+}$SR measurements on the General Purpose Surface-muon (GPS, temperature range $1.6$ K -- $150$ K) and Low Temperature Facility (LTF, temperature range $20$ mK -- $8$ K) spectrometers on the $\pi$M$3$ beamline of the S$\mu$S muon source of the Paul Scherrer Institute (PSI), Switzerland. All the measurements were performed under conditions of zero magnetic field (ZF). The samples were loose or pressed powders wrapped in Al/Mylar tape (GPS) or pressed powders glued onto a silver plate with Apiezon N grease and covered with a 12-$\mu$m-thick silver foil (LTF).

\begin{figure*}[t!]
	\vspace{3.6cm} \includegraphics{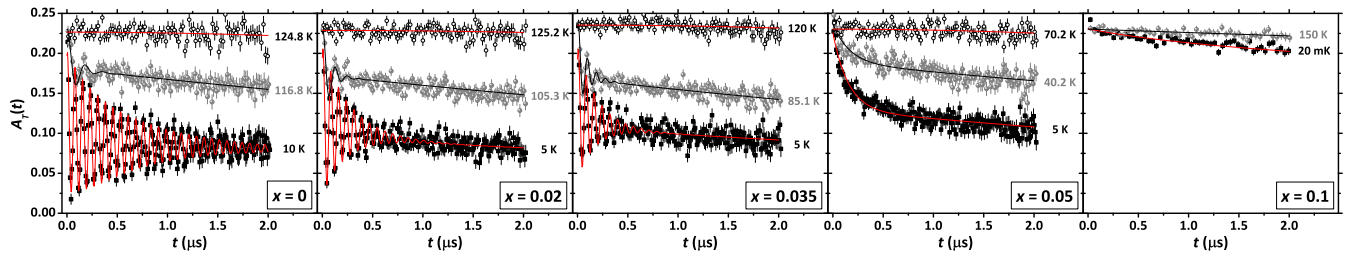}
	\caption{\label{GraMuSR} Representative time-domain $\mu^{+}$ spin depolarization curves for Eu$_{2}$Ir$_{2}$O$_{7}$ and for the four investigated $\left(\textrm{Eu}_{1-x}\textrm{Bi}_{x}\right)_{2}$Ir$_{2}$O$_{7}$ compounds at ambient pressure (ZF conditions). The continuous lines are best-fitting curves according to the detailed analysis reported in the main text.}
\end{figure*}
We also performed measurements \cite{Pra16} on Eu$_{2}$Ir$_{2}$O$_{7}$ powders on the General Purpose Decay-channel (GPD, temperature range $5$ K $\leq T \leq 200$ K, low-temperature pressure range $1$ bar $\leq P \leq 24$ kbar) spectrometer on the $\mu$E1 beamline of S$\mu$S at PSI, both in ZF conditions and while applying a weak transverse external magnetic field (TF). We applied pressure at ambient $T$ by means of a double-wall piston-cylinder cell made of MP$35$N alloy and ensured nearly-hydrostatic $P$ conditions in the whole experimental range by using Daphne oil $7373$ as transmitting medium. We determined the $P$ value and its homogeneity at low $T$ by measuring the diamagnetic response associated with the superconducting transition of a small In wire in the sample space by means of ac susceptibility.

Details on the synthesis and characterization of the samples can be found in Ref.~\cite{Tel19}. General technical details on $\mu^{+}$SR are reported in Ref.~\cite{SI} together with a discussion of the data analysis for the measurements under applied pressure.

\section{Results and data analysis}

In Fig.~\ref{GraMuSR} we report representative results of ZF-$\mu^{+}$SR for Eu$_{2}$Ir$_{2}$O$_{7}$ \cite{Pra16} and for the four investigated $\left(\textrm{Eu}_{1-x}\textrm{Bi}_{x}\right)_{2}$Ir$_{2}$O$_{7}$ samples ($x = 0.02, 0.035, 0.05, 0.1$). At low $T$, the experimental results for $x \leq 0.035$ evidence clear coherent oscillations of the asymmetry function $A_{T}(t)$ \cite{SI} -- denoting a well-defined value for $B_{\mu}$, i.e., the local magnetic field at the muon site -- while the data for $x = 0.05$ evidence a strong depolarization in the absence of coherent oscillations, indicative of a wide distribution of local $B_{\mu}$ values. Eventually, we detect no significant dependence of the muon spin polarization on temperature for $x = 0.1$, suggesting a full suppression of the magnetic phase in that limit of chemical dilution.

In order to reach a comprehensive picture for these results, we refer to the fitting function
\begin{eqnarray}\label{EqGeneralFittingZF}
\frac{A_{T}(t)}{A_{T}(0)} &=& \left[1 - V_{\textrm{m}}(T)\right] e^{-\frac{\sigma_{\textrm{N}}^{2} t^{2}}{2}} {}\\ && + \sum_{i=1}^{N} a_{i}^{\perp}(T) F_{i}(t) D_{i}^{\perp}(t) + a^{\parallel}(T) D^{\parallel}(t),\nonumber
\end{eqnarray}
which is generally used for magnetic materials in ZF conditions \cite{Pra16}. In Eq.~\eqref{EqGeneralFittingZF}, $V_{\textrm{m}}(T)$ is the fraction of muons probing a static local magnetic field and, due to the macroscopically-random implantation of muons, is equivalent to the magnetic volume fraction of the sample. In the high-$T$ paramagnetic phase, $V_{\textrm{m}}(T) = 0$ and only nuclear magnetic moments can cause a Gaussian-like damping of the signal with a characteristic rate $\sigma_{\textrm{N}} \sim 0.1 \; \mu$s$^{-1}$. In the low-$T$ magnetic phase, the superscript $\perp$ ($\parallel$) refers to those muons probing a static local magnetic field perpendicularly (parallelly) to the initial spin orientation. Accordingly, one has $\left[\sum_{i} a_{i}^{\perp}(T) + a^{\parallel}(T)\right] = V_{\textrm{m}}(T)$ for the so-called ``transverse'' ($a_{i}^{\perp}$ or $a_{i}^{Tr}$) and ``longitudinal'' ($a^{\parallel}$ or $a^{L}$) amplitudes. The index $i$ runs over $N$ inequivalent crystallographic implantation sites. The transverse component yields information about the static magnetic properties of the investigated phase. In particular, a precession of the muon spin around a local magnetic field static within the $\mu^{+}$ lifetime can be observed in the transverse amplitude and described by oscillating functions $F_{i}(t)$, while the damping functions $D_{i}^{\perp}(t)$ reflect a distribution of local magnetic fields at the $\mu^{+}$ site. On the other hand, the longitudinal components typically probe dynamical spin-lattice-like relaxation processes resulting in slow exponentially-decaying functions $D_{i}^{\parallel}(t) = e^{-\lambda_{i}^{L}t}$. Due to the typically low values measured for $\lambda_{i}^{L}$ ($\sim 0.1$ $\mu$s$^{-1}$) in comparison with the overall experimental $t$-window ($\sim 5$ $\mu$s), the $i$ different longitudinal components cannot be resolved and only one average $D^{\parallel}(t)$ is reported in Eq.~\eqref{EqGeneralFittingZF}, accordingly.

In the following, we define $T_{N}$ as the temperature where a crossover takes place between $a^{\perp} = 0$ and $a^{\perp} \neq 0$. This definition is in good agreement (within $1$ -- $2$ K degrees) with the estimates from magnetometry \cite{Tel19}.
\begin{figure*}[t!]
	\vspace{10.6cm} \includegraphics{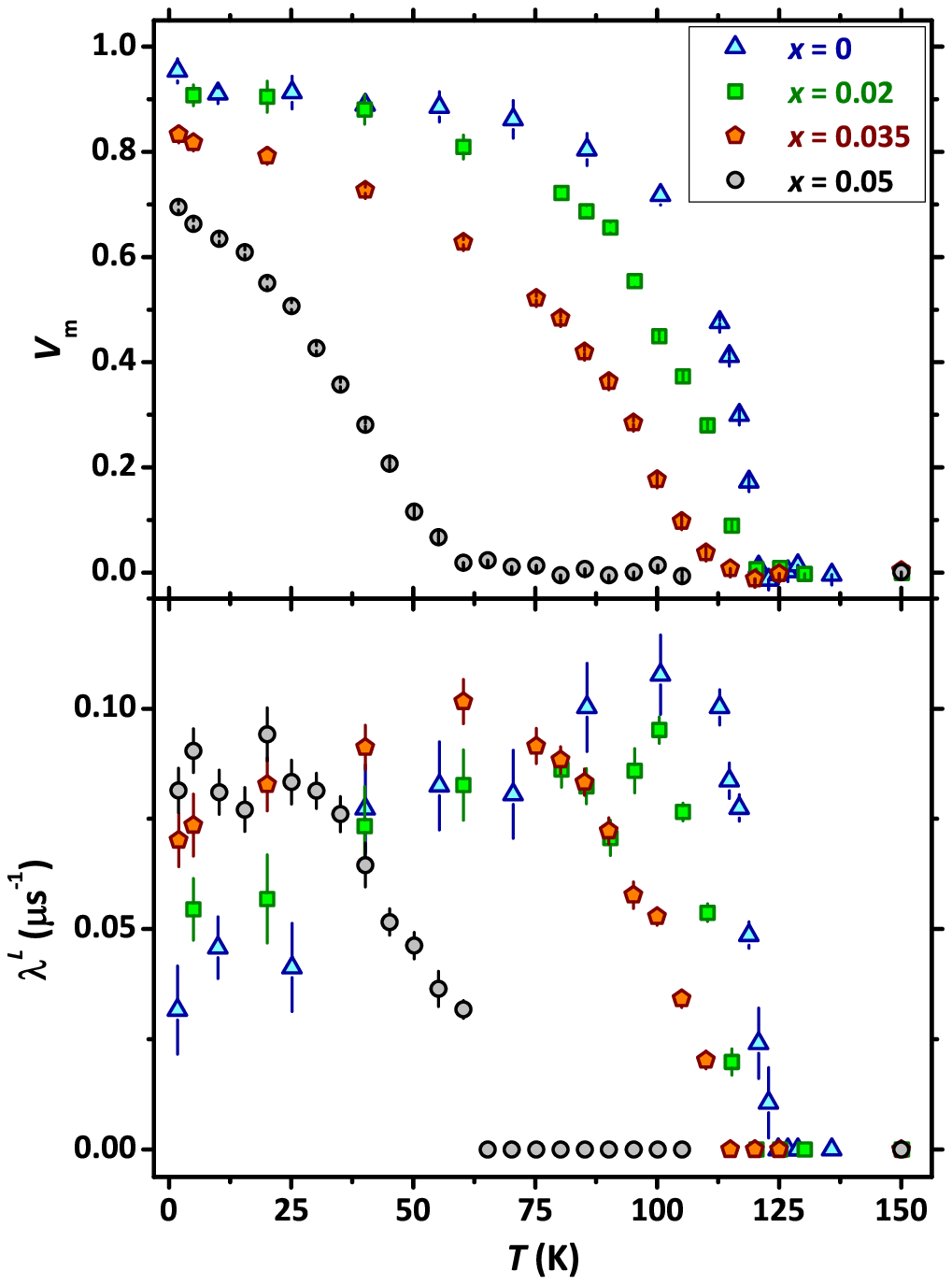} \includegraphics{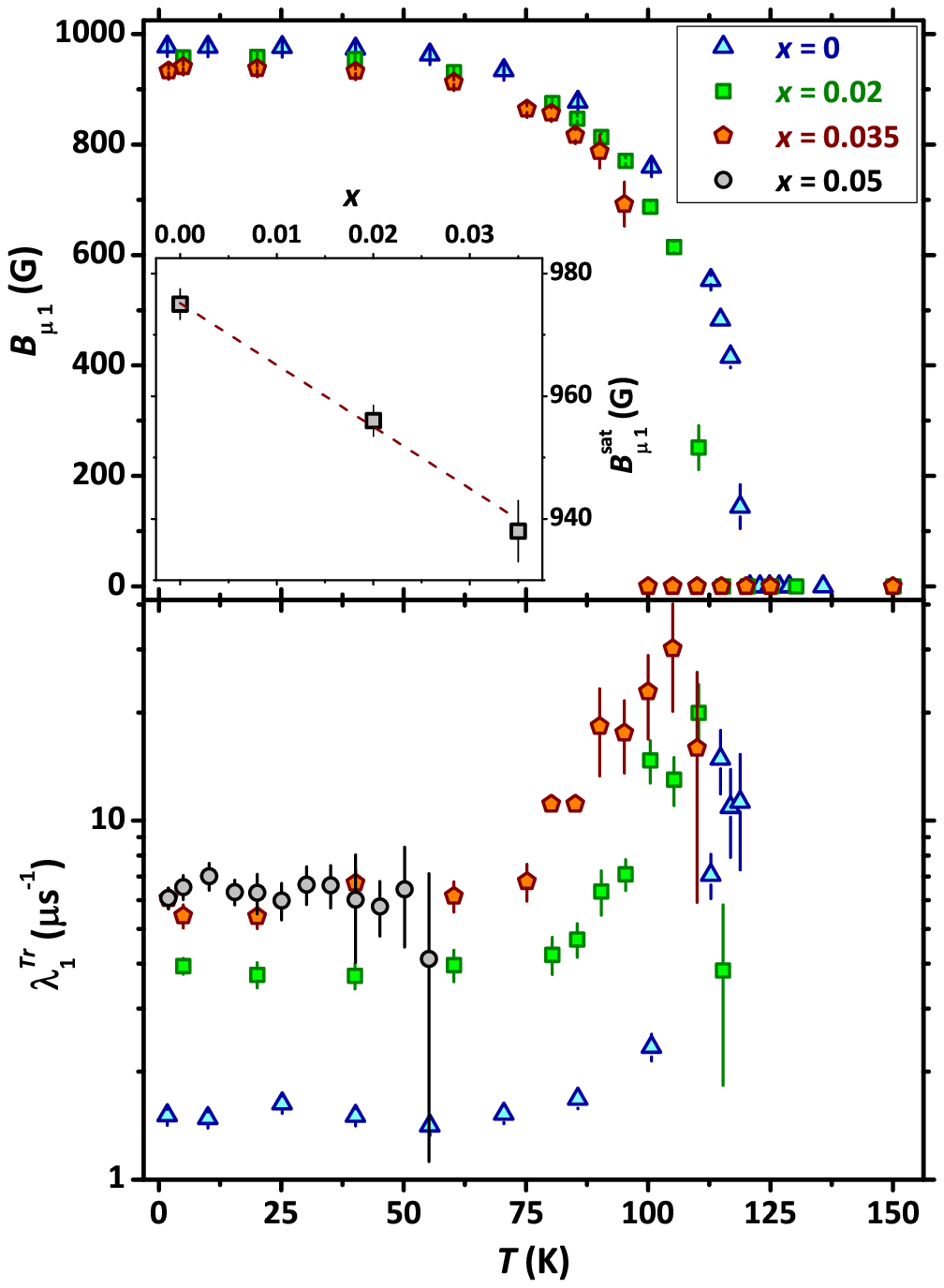}
	\caption{\label{GraVm} Left-hand panel: temperature dependence of the magnetic volume fraction (upper panel) and of the longitudinal relaxation (lower panel) for $\left(\textrm{Eu}_{1-x}\textrm{Bi}_{x}\right)_{2}$Ir$_{2}$O$_{7}$ ($x \leq 0.05$) at ambient pressure. Right-hand panel: temperature dependence of the internal field corresponding to the coherent oscillations (upper panel, main) and of the corresponding transverse damping rate (lower panel) for $\left(\textrm{Eu}_{1-x}\textrm{Bi}_{x}\right)_{2}$Ir$_{2}$O$_{7}$ ($x \leq 0.035$) at ambient pressure. Upper panel, inset: low-temperature saturation value for the internal field as a function of the Bi content. The dashed line is a guide to the eye.}
\end{figure*}

\subsubsection{Parent compound Eu$_{2}$Ir$_{2}$O$_{7}$ at ambient pressure}

We discussed the results for the Eu$_{2}$Ir$_{2}$O$_{7}$ sample in detail in a previous work (see the supplementary material of Ref.~\onlinecite{Pra16}) and we will briefly summarize them here.

With reference to Eq.~\eqref{EqGeneralFittingZF}, our experimental results suggest that the muons are implanted in $N = 2$ inequivalent crystallographic sites and that $a_{1}^{\perp}/a_{2}^{\perp} \sim 3.5 - 4$ independently on $T$. The main component $a_{1}^{\perp}$ displays long-lived coherent oscillations indicative of a well-defined value of the local magnetic field with a narrow distribution over the sample. We find good fitting results with the choice $F_{1}(t) = \cos\left(\gamma B_{\mu 1} t + \phi_{1}\right)$ and $D_{1}^{\perp}(t) = e^{-\lambda_{1}^{Tr}t}$. A preliminary fitting procedure shows that the phase term $\phi_{1} \sim -20$°, so that the final fitting procedure was performed by keeping $\phi_{1} = -20$° as fixed parameter. $a_{2}^{\perp}$ is a non-oscillating component and, accordingly, $F_{2}(t) = 1$ with the only contribution to the depolarization coming from $D_{2}^{\perp}(t) = e^{-\lambda_{2}^{Tr}t}$, where $\lambda_{2}^{Tr} \sim 4 \; \mu$s$^{-1}$. Based on the criterion above, we estimate $T_{N} = 120.0 \pm 1.0$ K.

\subsubsection{$x = 0.02$ and $x = 0.035$ samples at ambient pressure}

The fitting procedure still suggests that $a_{1}^{\perp}/a_{2}^{\perp} \sim 3.5 - 4$ holds roughly independent on $T$ for the signals from the two inequivalent crystallographic sites. The long-lived coherent oscillations are fitted properly by $F_{1}(t) = \cos\left(\gamma B_{\mu 1} t + \phi_{1}\right)$ and $D_{1}^{\perp}(t) = e^{-\lambda_{1}^{Tr}t}$, similarly to the case of Eu$_{2}$Ir$_{2}$O$_{7}$. The best fitting results are obtained by using $\phi_{1} = -32$° and $\phi_{1} = -38$° as fixed parameters for $x = 0.02$ and $x = 0.035$, respectively. We stress that in a narrow temperature window just below $T_{N}$ the transverse relaxation is so strong that the coherent oscillations are overdamped and $B_{\mu 1}$ cannot be defined. Based on the criterion above, we estimate $T_{N} = 118.0 \pm 2.5$ K and $T_{N} = 112.5 \pm 2.5$ K for $x = 0.02$ and $x = 0.035$, respectively.

\subsubsection{$x = 0.05$ and $x = 0.1$ samples at ambient pressure}

As shown in Fig.~\ref{GraMuSR}, the coherent oscillations observed for $x \leq 0.035$ are completely overdamped for $x = 0.05$ at all the investigated $T$ values, although for this latter composition the results still evidence a strongly $T$-dependent asymmetry function. Accordingly, we impose $F_{1}(t) = 1$ and associate the whole transverse depolarization to the function $D_{1}^{\perp}(t) = e^{-\lambda_{1}^{Tr}t}$. Moreover, the fitting quality is satisfactory even without including the second transverse component -- accordingly, we set $a_{2}^{\perp} = 0$. We estimate $T_{N} = 57.5 \pm 2.5$ K.

Finally, the $x = 0.1$ sample does not evidence clear signs of magnetism down to $20$ mK, although we observe a clear difference in the relaxation shape between the measurements at the lowest and highest temperatures. We ascribe the origin of the low-temperature relaxation to extrinsic magnetic impurities, whose contribution to the macroscopic magnetization was already evidenced for $T \lesssim 10$ K \cite{Tel19}. Accordingly, we assume $T_{N} < 20$ mK.

\section{Discussion}

We report the $T$-dependence of the magnetic volume fraction for $x \leq 0.05$ in the upper panel of Fig.~\ref{GraVm} (left-hand panel). The onset of the increase of $V_{m}$ in Eu$_{2}$Ir$_{2}$O$_{7}$ is very sharp at $T_{N}$ and the saturation value for $V_{m}$ at low temperatures is around $0.9$, in agreement with previous observations \cite{Zha11}. We associate the remaining fraction $\sim 0.1$ to a segregated non-magnetic phase and/or to a non-relaxing component arising from those muons implanted in the cryostat walls and sample holder. We also observe a sharp increase in the longitudinal relaxation $\lambda^{L}$ at $T_{N}$, followed by an unusually slow decrease with further cooling within the magnetic phase (see the lower panel of Fig.~\ref{GraVm}, left-hand panel). These results are indicative of a canonical dynamical critical peak at $T_{N}$ -- associated with the slowing down of magnetic fluctuations upon approaching the critical point -- and of the persistence of more unconventional spin dynamics within the magnetic phase, possibly associated with the random movement of the magnetic domain walls \cite{Yam14,Ma15,Tar15,Hir17,Kim18}.

The same results are observed also for the $x = 0.02$ sample, although the increase of $V_{m}$ upon decreasing $T$ is slower than what is observed for Eu$_{2}$Ir$_{2}$O$_{7}$. For the $x = 0.035$ sample we observe an overall shift to lower temperatures, consistent with the decrease of $T_{N}$, and a rounding of the $V_{m}$ onset. Additionally, the low-temperature saturation value for $V_{m}$ is around $0.8$, denoting the increase of the segregated non-magnetic volume. We confirm this trend for the $x = 0.05$ sample, where the shift of the $V_{m}$ onset to lower temperatures is much more marked and where $V_{m} \sim 0.7$ at saturation. These results are paralleled in the behaviour of $\lambda^{L}$, also confirming the unusual persistence of spin dynamics within the magnetic phase for all the samples $x \leq 0.05$.

Generally, from the $\mu^{+}$SR perspective, a magnetic phase is LRO (long-range ordered) if the distribution of local magnetic fields around the average value is narrow enough (i.e., $B_{\mu 1} \gg \lambda_{1}^{Tr}/\gamma$). In this case, $B_{\mu 1}$ can be considered as the order parameter for the studied phase transition. On the other hand, in a SRO (short-range ordered) phase the magnetic response is robust and extended over a bulk fraction of the sample -- however, the field distribution is so broad that it is not meaningful to define a central average value and, correspondingly, no coherent oscillations are observed in the experimental $A_{T}(t)$ asymmetry curves. Both Eu$_{2}$Ir$_{2}$O$_{7}$ and the $x = 0.02, 0.035$ samples belong to the former category and we plot the temperature dependence of their average $B_{\mu 1}$ values in the upper panel of Fig.~\ref{GraVm} (right-hand panel). The overall trends are qualitatively very similar for the three samples. Consistently with the results discussed above for $V_{m}$ and $\lambda^{L}$, it is evident that the effect of increasing $x$ is to weaken the magnetic phase by suppressing both $T_{N}$ and $B_{\mu 1}^{\textrm{sat}}$, i.e., the low-temperature saturation value of the internal field. This latter quantity is reported in the inset, clearly showing a linear-like decrease upon increasing $x$. On the other hand, a complete crossover to a SRO state is observed for the $x = 0.05$ sample. This is supported by the lower panel of Fig.~\ref{GraVm} (right-hand panel), where we report the $T$-dependence of the transverse damping rate $\lambda_{1}^{Tr}$. It is evident that the low-temperature value of $\lambda_{1}^{Tr}$ increases progressively upon increasing $x$ which, together with the observed $x$-induced suppression of $B_{\mu 1}^{\textrm{sat}}$, eventually leads to the complete overdamping of the coherent oscillations.

\subsection{Unusual behaviour for the magnetic volume fraction of Eu$_{\bf{2}}$Ir$_{\bf{2}}$O$_{\bf{7}}$}

\begin{figure*}[t!]
	\vspace{15.2cm} \includegraphics{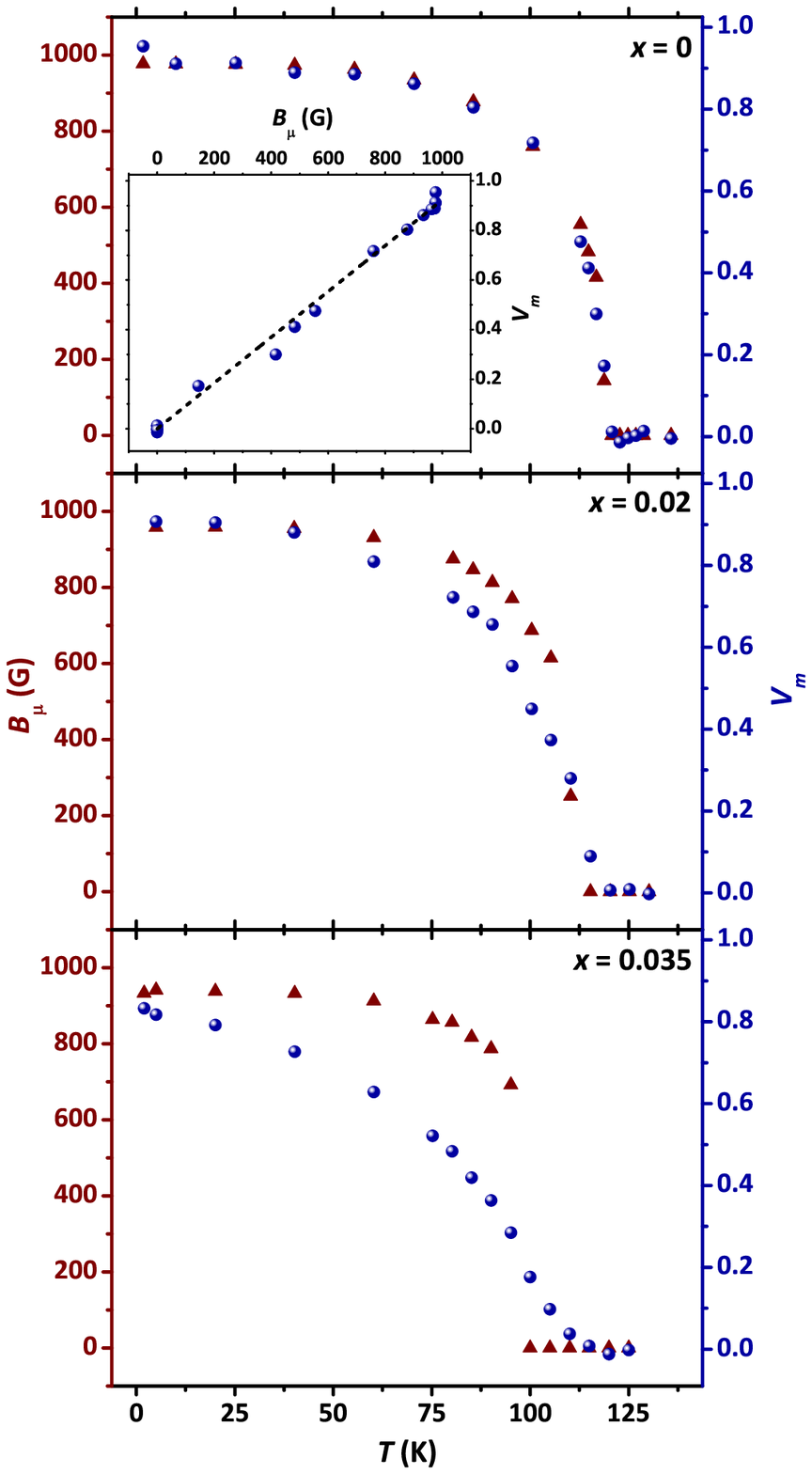} \includegraphics{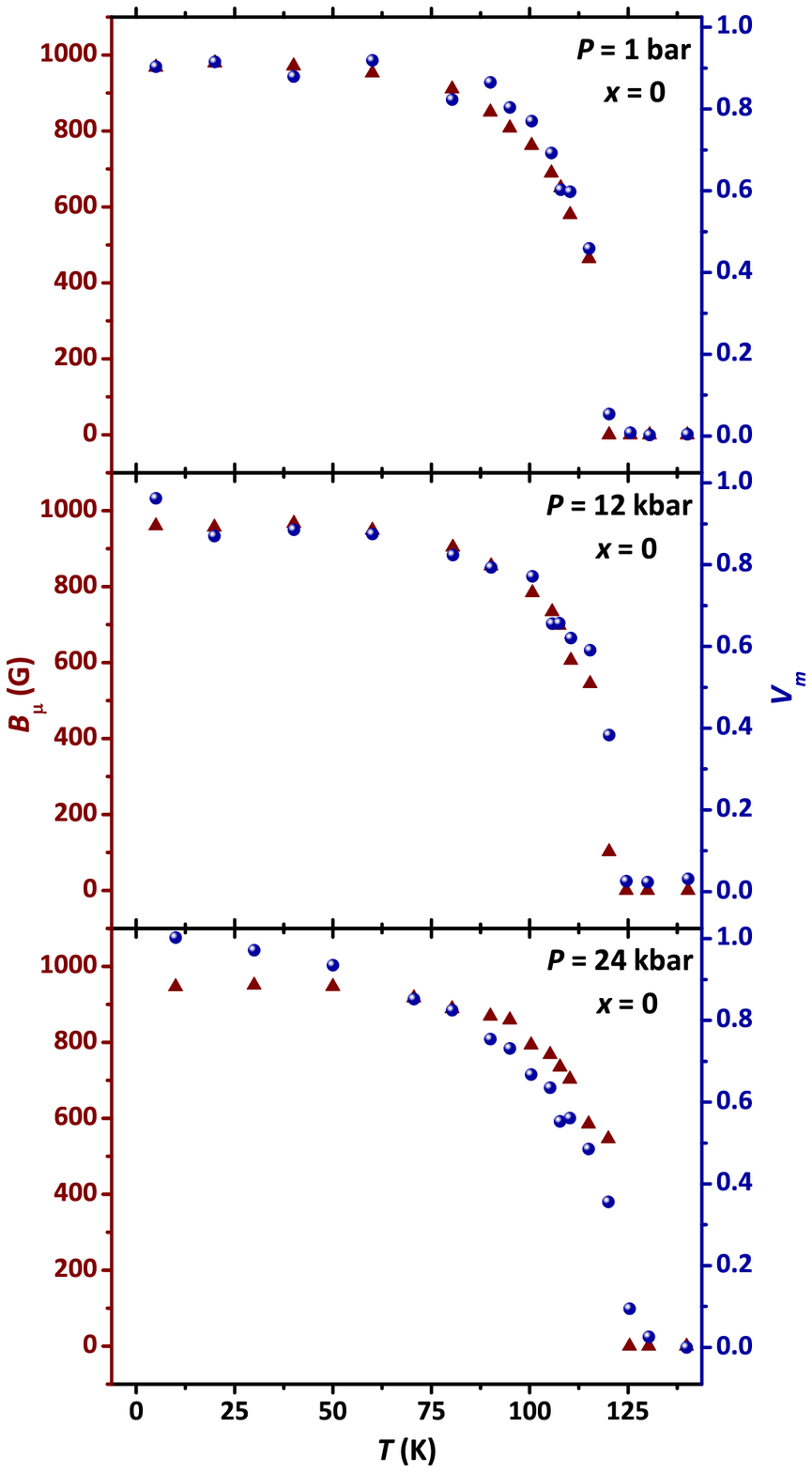}
	\caption{\label{GraBiVsP}(Color online) Evolution of $B_{\mu}(T)$ and $V_{m}(T)$ for Eu$_{2}$Ir$_{2}$O$_{7}$ as a function of different perturbations. Left-hand panel: effect of the Eu$_{1-x}$Bi$_{x}$ substitution in the low-$x$ regime obtained from ZF $\mu^{+}$SR. For the undoped material, the same data are also reported in the inset with $V_{m}$ displayed as a function of $B_{\mu}$ and temperature as implicit parameter. The dashed curve is a best-fitting linear trend with zero-intercept. Right-hand panel: effect of the external pressure estimated from weak-transverse-field $\mu^{+}$SR \cite{Pra16,SI}. The graph corresponding to $P = 1$ bar is physically equivalent to the upper left-hand panel, however the data have been obtained for the sample inside the unloaded pressure cell so that the signal-to-noise ratio is worse \cite{SI}.}
\end{figure*}
In magnetic materials, the possibility of quantifying separately the order parameter and the volume fraction over which the ordered phase extends is a powerful advantage of $\mu^{+}$SR \cite{Blu99,Yao11}. The former quantity -- i.e., the ordered magnetic moment -- is proportional to the local magnetic field at the $\mu^{+}$ site, hereafter indicated as $B_{\mu}$. The proportionality factor can be estimated by means of probabilistic approaches \cite{Blu12,Dis14} or computed exactly once the $\mu^{+}$ thermalization site and interaction mechanisms are known \cite{Mol13,Bon16}. However, the temperature dependence for $B_{\mu}$ and for the order parameter is the same, which is relevant in order to estimate experimentally the associated critical exponent $\beta$ for $T \lesssim T_{c}$ -- where $T_{c}$ represents the broadly defined critical temperature \cite{Sta71}. On the other hand, thanks to the macroscopically-random implantation of muons during the experiment, the magnetic volume fraction of the sample $V_{m}$ can be quantified as the fraction of muons probing a static local magnetic field within the muon lifetime.

The dependences of $B_{\mu}$ and $V_{m}$ on temperature are expected to be different, as confirmed by experiments routinely \cite{Kla08,Pra13,Mat15,Fra16,Wil16}. Let's focus on an idealized, textbook-like phase transition involving identical localized magnetic moments on a lattice. For continuous phase transitions, $B_{\mu}$ follows a power-law trend governed by the characteristic critical exponent $\beta$ for $T \lesssim T_{c}$. For $T > T_{c}$ the order parameter is zero by definition so the overall behaviour is \textit{intrinsically non-symmetric} around $T_{c}$. On the other hand, based on naive Landau-like considerations on the minimization of the free energy \cite{Cha95}, a magnetic phase is suddenly favoured over the whole sample volume below $T_{c}$ and $V_{m}(T)$ should follow an inverted step-like function centred at $T_{c}$. A supposedly normal distribution of $T_{c}$ values with characteristic width $\Delta$ -- due to microscopic inhomogeneities -- leads to the expression \cite{Pra13}
\begin{equation}\label{EqMagneVolERFC}
V_{m}(T) = \frac{1}{2} \; \textrm{erfc}\left[\frac{T - T_{c}} {\sqrt{2}\Delta}\right],
\end{equation}
involving the complementary error function $\textrm{erfc}(x)$ and being routinely used as best-fitting function for experimental $V_{m}(T)$ data. According to Eq.~\eqref{EqMagneVolERFC}, the function $V_{m}(T)$ is \textit{symmetric} -- in particular, it is odd with respect to the mid-point $T_{c}$, which is then defined as an average value.

Our experimental findings for Eu$_{2}$Ir$_{2}$O$_{7}$ are at variance with the expected behaviours outlined above. As reported in the upper left-hand panel of Fig.~\ref{GraBiVsP}, we evidence a very peculiar correlation between $B_{\mu}(T)$ and $V_{m}(T)$. In particular, the functional form describing the dependence of both quantities on $T$ looks identical, as shown in the inset where $V_{m}$ is reported as a function of $B_{\mu}$ with $T$ as implicit parameter. It is crucial to stress that $B_{\mu}$ is obtained as best-fitting parameter for the frequency of the coherent oscillations in the transverse component of the $\mu^{+}$SR depolarization function, while $V_{m}$ is derived from the amplitude of the non-oscillating longitudinal component of the signal. Accordingly, it seems safe to assume that these two quantities are estimated independently from one another, as there is no immediate way of understanding any numerical artefact of the fitting procedure leading to a correlation between them.

It is interesting to discuss the effect of different perturbations on the observed correlation between $B_{\mu}$ and $V_{m}$ in Eu$_{2}$Ir$_{2}$O$_{7}$. In the left-hand panels of Fig.~\ref{GraBiVsP} we report data for Eu$_{2}$Ir$_{2}$O$_{7}$ under the effect of the Eu$_{1-x}$Bi$_{x}$ chemical substitution in the low-$x$ regime. It is evident that the correlation between $B_{\mu}(T)$ and $V_{m}(T)$ is affected already for $x = 0.02$ and that the deviations become substantial for $x = 0.035$, leading to a result which resembles much more the expected canonical trend. Our $\mu^{+}$SR measurements detect a clear crossover to a magnetically-disordered state for $x = 0.05$ and a complete suppression of magnetism for $x = 0.1$. Accordingly, it is clear that the effect of low $x$ values is to destabilize magnetism and, in particular, the all-in--all-out state. The situation is different by considering the effect of nearly-hydrostatic external pressure for values $P \leq 24$ kbar (Fig.~\ref{GraBiVsP}, right-hand panels). The peculiar correlation between $B_{\mu}(T)$ and $V_{m}(T)$ is preserved up to $P = 24$ kbar, even though we detect minor departures from it at the highest measured pressure value. Overall, the preservation of this phenomenology is consistent with the weak sensitivity of the all-in--all-out phase and of the Ir$^{4+}$ magnetic moment to pressure reported previously \cite{Pra16}.

Based on these observations, we argue -- as the main result of this work -- that the observed correlation between $B_{\mu}(T)$ and $V_{m}(T)$ is a peculiar property of the all-in--all-out phase itself, as realized in stoichiometric Eu$_{2}$Ir$_{2}$O$_{7}$.

\subsection{Magnetic droplets. Hedgehog monopoles and their possible role in limiting the growth of the magnetic volume}

It is clear that, in Eu$_{2}$Ir$_{2}$O$_{7}$, the temperature dependence of $V_{m}$ -- rather than $B_{\mu}$ -- defies the usual expectations. Accordingly, the main assumption for an interpretation of the observed phenomenology should consider the development of an ordered magnetic moment at the Ir site as the primary parameter for the magnetic transition, the magnetic volume fraction being effectively driven by the order parameter.

In particular, the process can be pictured by assuming that a local non-zero ordered magnetic moment nucleates within the sample at different point-like regions at $T_{N}$ generating a magnetically-correlated volume proportional to the amplitude of the ordered moment. The nature of these nucleation points is currently unknown, however it seems safe to assume that structural or chemical defects may play this role. The peculiar shape of $V_{m}(T)$ suggests that the volume expansion of the magnetic droplets occurs within a non-magnetic background. It is important to stress that $\mu^{+}$SR is resolving a well-defined local magnetic field already for $T \lesssim T_{N}$. This fact means that the magnetic phase is long-range ordered beyond at least tens of lattice parameters from the $\mu^{+}$ implantation site. Accordingly, the nucleated regions must be expanding quickly in volume upon decreasing temperature below $T_{N}$ and, crucially, a long-range ordered phase must be established within the magnetic droplets. A further decrease of temperature would progressively increase the amplitude of the ordered moment and this would polarize a bigger ordered volume around the nucleation point, in turn. The finding of a low-temperature saturation value $V_{m} \sim 90$ \% is also confirmed by other works \cite{Zha11} and could be an indication that the expansion of the nucleated droplets continues until they jam together, leaving small non-magnetic interstitial voids between them.

\begin{figure}[t!]
	\vspace{7.6cm} \includegraphics{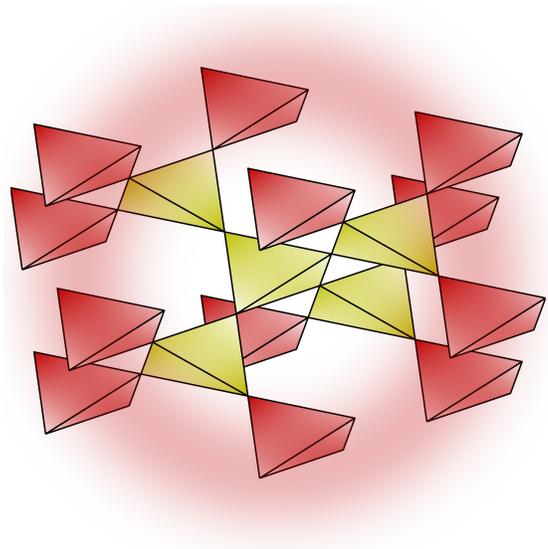}
	\caption{\label{GraNucleation}(Color online) During the expansion of the magnetically-ordered volume, represented by the yellow tetrahedra, new hedgehog monopoles must be generated corresponding to the red tetrahedra on the surface of the droplet. The associated energy cost acts as a limiting factor for the growth of small droplets.}
\end{figure}
The described phenomenology is reminiscent of the condensation of vapour \cite{Abr74,Blu06}. In such framework, the growth of a condensed spherical droplet of liquid is governed by two competing terms in the overall energy balance. In particular, a negative term accounting for the volume energy gain of the system in condensing vapour into liquid is opposed by a positive term associated with the surface tension. Naively, two analogous competing energy terms can be considered when modelling the magnetic phase transition of Eu$_{2}$Ir$_{2}$O$_{7}$. The conventional volume energy gain below the critical temperature needs to be opposed by a positive energy contribution from the surface. We argue that this latter term should be associated with the non-trivial topological properties of the all-in--all-out phase. The arrangement of the magnetic moments within each tetrahedron in the ordered phase can be described as a hedgehog magnetic monopole \cite{Ari13,Bra12,Zan18}. Accordingly, the crossover between paramagnetic and ordered states implies a simultaneous crossover between topologically trivial and non-trivial states -- respectively -- for each tetrahedron. The energy cost of the generation of new hedgehog monopoles should play as a limiting factor for the growth of the magnetic volume of the ordered phase. This is represented in Fig.~\ref{GraNucleation} for the idealized case of a spherical droplet of ordered tetrahedra (in yellow) surrounded by those tetrahedra (in red) that must be converted into hedgehog magnetic monopoles during the expansion process. The number of new generated hedgehog monopoles -- and the associated energy contribution -- clearly scales with the surface of the droplet. Upon increasing the droplet size, the energy gain from the volume term is expected to dominate over the surface term.

The analogy with vapour condensation should be used with care. Nucleation is normally associated with first-order-like transitions where metastable states emerge -- e.g., the supercooled gas. Current experimental reports have failed to identify the magnetic transition in Eu$_{2}$Ir$_{2}$O$_{7}$ as first-order \cite{Ish12}. Additionally, our results of dc magnetometry (not shown) do not evidence any dependence of $T_{N}$ upon cooling the sample at different rates from $1$ K/min to $20$ K/min. Still, the highly-unusual findings reported above point at least towards a weak first-order character of the transition and deserve further investigations. In particular, it would be highly desirable to use a detailed Landau-like expansion for the characteristic free energy of Eu$_{2}$Ir$_{2}$O$_{7}$ in the magnetic state -- including the proposed energy cost for the creation of new hedgehog monopoles -- aiming at the computational simulation of the expansion of the magnetic volume upon decreasing temperature.

\section{Conclusions}

In this paper we have presented a detailed study of the magnetic phase transition of Eu$_{2}$Ir$_{2}$O$_{7}$ by means of muon-spin spectroscopy. The growth of the magnetic volume upon decreasing temperature is unusually slow and it is possibly limited by the non-trivial topological properties of the magnetically-ordered tetrahedra, supporting the development of an all-in/all-out arrangement for the magnetic moments of Ir ions. Our data on Eu$_{2}$Ir$_{2}$O$_{7}$ under external pressure and, at the same time, under the effect of the Eu$_{1-x}$Bi$_{x}$ chemical substitution show that the proposed mechanism is highly specific for the undoped stoichiometric composition. The novel exotic mechanism proposed for the nucleation of the magnetic volume deserves confirmation by additional experimental and computational investigations.

\section*{Acknowledgements}

G. P. acknowledges the kind support of the Laboratory for Muon Spin Spectroscopy at the Paul Scherrer Institute during the $\mu^{+}$SR experiments and stimulating discussions with Seung-Ho Baek, Pietro Carretta and Jos\'e Lorenzana. S. D. W. acknowledges support from the NSF award DMR-1905801.




\begin{thebibliography}{99}
	\bibitem{Gar10} J. S. Gardner, M. J. P. Gingras, J. E. Greedan, {\it Magnetic pyrochlore oxides}, Rev. Mod. Phys. {\bf 82} 53 (2010).
	\bibitem{Shi19} H. Shinaoka, Y. Motome, T. Miyake, S. Ishibashi, P. Werner, {\it First-principles studies of spin-orbital physics in pyrochlore oxides}, Journ. Phys.: Cond. Matt. {\bf 31} 323001 (2019).
	\bibitem{Pes10} D. Pesin, L. Balents, {\it Mott physics and band topology in materials with strong spin-orbit interaction}, Nat. Phys. {\bf 6} 376 (2010).
	\bibitem{Wan11} X. Wan, A. M. Turner, A. Vishwanath, S. Y. Savrasov, {\it Topological semimetal and Fermi-arc surface states in the electronic structure of pyrochlore iridates}, Phys. Rev. B {\bf 83} 205101 (2011).
	\bibitem{Moo13} E.-G. Moon, C. Xu, Y. B. Kim, L. Balents, {\it Non-Fermi-Liquid and Topological States with Strong Spin-Orbit Coupling}, Phys. Rev. Lett. {\bf 111} 206401 (2013).
	\bibitem{Wit14} W. Witczak-Krempa, G. Chen, Y. B. Kim, L. Balents, {\it Correlated Quantum Phenomena in the Strong Spin-Orbit Regime}, Annu. Rev. Condens. Matter Phys. {\bf 5} 57 (2014).
	\bibitem{Sus15} A. B. Sushkov, J. B. Hofmann, G. S. Jenkins, J. Ishikawa, S. Nakatsuji, S. Das Sarma, H. D. Drew, {\it Optical evidence for a Weyl semimetal state in pyrochlore Eu$_{2}$Ir$_{2}$O$_{7}$}, Phys. Rev. B {\bf 92} 241108(R) (2015).
	\bibitem{Gos17} P. Goswami, B. Roy, S. Das Sarma, {\it Competing orders and topology in the global phase diagram of pyrochlore iridates}, Phys. Rev. B {\bf 95} 085120 (2017).
	\bibitem{Ber18} C. Berke, P. Michetti, C. Timm, {\it Stability of the Weyl-semimetal phase on the pyrochlore lattice}, New Journ. Phys. {\bf 20} 043057 (2018).
	\bibitem{Tai01} N. Taira, M. Wakeshima, Y. Hinatsu, {\it Magnetic properties of iridium pyrochlores $R_{2}$Ir$_{2}$O$_{7}$ ($R$ = Y, Sm, Eu and Lu)}, Journ. Phys.: Cond. Matt. {\bf 13} 5527 (2001).
	\bibitem{Yan01} D. Yanagishima, Y. Maeno, {\it Metal-Nonmetal Changeover in Pyrochlore Iridates}, Journ. Phys. Soc. Jpn. {\bf 70} 2880 (2001).
	\bibitem{Mat07} K. Matsuhira, M. Wakeshima, R. Nakanishi, T. Yamada, A. Nakamura, W. Kawano, S. Takagi, Y. Hinatsu, {\it Metal-Insulator Transition in Pyrochlore Iridates $Ln_{2}$Ir$_{2}$O$_{7}$ ($Ln =$ Nd, Sm, and Eu)}, Journ. Phys. Soc. Jpn. {\bf 76} 043706 (2007).
	\bibitem{Mat11} K. Matsuhira, M. Wakeshima, Y. Hinatsu, S. Takagi, {\it Metal-Insulator Transitions in Pyrochlore Oxides $Ln_{2}$Ir$_{2}$O$_{7}$}, Journ. Phys. Soc. Jpn. {\bf 80} 094701 (2011).
	\bibitem{Zha11} S. Zhao, J. M. Mackie, D. E. MacLaughlin, O. O. Bernal, J. J. Ishikawa, Y. Ohta, and S. Nakatsuji, {\it Magnetic transition, long-range order, and moment fluctuations in the pyrochlore iridate Eu$_{2}$Ir$_{2}$O$_{7}$}, Phys. Rev. B {\bf 83} 180402(R) (2011).
	\bibitem{Dis12} S. M. Disseler, C. Dhital, T. C. Hogan, A. Amato, S. R. Giblin, C. de la Cruz, A. Daoud-Aladine, S. D. Wilson, M. J. Graf, {\it Magnetic order and the electronic ground state in the pyrochlore iridate Nd$_{2}$Ir$_{2}$O$_{7}$}, Phys. Rev. B {\bf 85}, 174441 (2012).
	\bibitem{Ish12} J. J. Ishikawa, E. C. T. O'Farrell, S. Nakatsuji, {\it Continuous transition between antiferromagnetic insulator and paramagnetic metal in the pyrochlore iridate Eu$_{2}$Ir$_{2}$O$_{7}$}, Phys. Rev. B {\bf 85} 245109 (2012).
	\bibitem{Tom12} K. Tomiyasu, K. Matsuhira, K. Iwasa, M. Watahiki, S. Takagi, M. Wakeshima, Y. Hinatsu, M. Yokoyama, K. Ohoyama, K. Yamada, {\it Emergence of Magnetic Long-range Order in Frustrated Pyrochlore Nd$_{2}$Ir$_{2}$O$_{7}$ with Metal-Insulator Transition}, Journ. Phys. Soc. Jpn. {\bf 81} 034709 (2012).
	\bibitem{Dis13} S. M. Disseler, S. R. Giblin, C. Dhital, K. C. Lukas, S. D. Wilson, M. J. Graf, {\it Magnetization and Hall effect studies on the pyrochlore iridate Nd$_{2}$Ir$_{2}$O$_{7}$}, Phys. Rev. B {\bf 87} 060403(R) (2013).
	\bibitem{Guo13} H. Guo, K. Matsuhira, I. Kawasaki, M. Wakeshima, Y. Hinatsu, I. Watanabe, Z.-A. Xu, {\it Magnetic order in the pyrochlore iridate Nd$_{2}$Ir$_{2}$O$_{7}$ probed by muon spin relaxation}, Phys. Rev. B {\bf 88} 060411(R) (2013).
	\bibitem{Nak16} M. Nakayama, T. Kondo, Z. Tian, J. J. Ishikawa, M. Halim, C. Bareille, W. Malaeb, K. Kuroda, T. Tomita, S. Ideta, K. Tanaka, M. Matsunami, S. Kimura, N. Inami, K. Ono, H. Kumigashira, L. Balents, S. Nakatsuji, S. Shin, {\it Slater to Mott Crossover in the Metal to Insulator Transition of Nd$_{2}$Ir$_{2}$O$_{7}$}, Phys. Rev. Lett. {\bf 117} 056403 (2016).
	\bibitem{Asi17} R. Asih, N. Adam, S. S. Mohd-Tajudin, D. P. Sari, K. Matsuhira, H. Guo, M. Wakeshima, Y. Hinatsu, T. Nakano, Y. Nozue, S. Sulaiman, M. I. Mohamed-Ibrahim, P. K. Biswas, I. Watanabe, {\it Magnetic Moments and Ordered States in Pyrochlore Iridates Nd$_{2}$Ir$_{2}$O$_{7}$ and Sm$_{2}$Ir$_{2}$O$_{7}$ Studied by Muon-Spin Relaxation}, Journ. Phys. Soc. Jpn. {\bf 86} 024705 (2017).
	\bibitem{Zha17} H. Zhang, K. Haule, D. Vanderbilt, {\it Metal-Insulator Transition and Topological Properties of Pyrochlore Iridates}, Phys. Rev. Lett. {\bf 118} 026404 (2017).
	\bibitem{Nak06} S. Nakatsuji, Y. Machida, Y. Maeno, T. Tayama, T. Sakakibara, J. van Duijn, L. Balicas, J. N. Millican, R. T. Macaluso, J. Y. Chan, {\it Metallic Spin-Liquid Behavior of the Geometrically Frustrated Kondo Lattice Pr$_{2}$Ir$_{2}$O$_{7}$}, Phys. Rev. Lett. {\bf 96} 087204 (2006).
	\bibitem{Mac07} Y. Machida, S. Nakatsuji, Y. Maeno, T. Tayama, T. Sakakibara, S. Onoda, {\it Unconventional Anomalous Hall Effect Enhanced by a Noncoplanar Spin Texture in the Frustrated Kondo Lattice Pr$_{2}$Ir$_{2}$O$_{7}$}, Phys. Rev. Lett. {\bf 98} 057203 (2007).
	\bibitem{Tok14} Y. Tokiwa, J. J. Ishikawa, S. Nakatsuji, P. Gegenwart, {\it Quantum criticality in a metallic spin liquid}, Nat. Mater. {\bf 13} 356 (2014).
	\bibitem{Sav14} L. Savary, E.-G. Moon, L. Balents, {\it New Type of Quantum Criticality in the Pyrochlore Iridates}, Phys. Rev. X {\bf 4} 041027 (2014).
	\bibitem{Ari13} T. Arima, {\it Time-Reversal Symmetry Breaking and Consequent Physical Responses Induced by All-In--All-Out Type Magnetic Order on the Pyrochlore Lattice}, Journ. Phys. Soc. Jpn. {\bf 82} 013705 (2013).	
	\bibitem{Sag13} H. Sagayama, D. Uematsu, T. Arima, K. Sugimoto, J. J. Ishikawa, E. O'Farrell, S. Nakatsuji, {\it Determination of long-range all-in--all-out ordering of Ir$^{4+}$ moments in a pyrochlore iridate Eu$_{2}$Ir$_{2}$O$_{7}$ by resonant x-ray diffraction}, Phys. Rev. B {\bf 87} 100403(R) (2013).	
	\bibitem{Dis14} S. M. Disseler, {\it Direct evidence for the all-in/all-out magnetic structure in the pyrochlore iridates from muon spin relaxation}, Phys. Rev. B {\bf 89} 140413(R) (2014).
	\bibitem{Don16} C. Donnerer, M. C. Rahn, M. Moretti Sala, J. G. Vale, D. Pincini, J. Strempfer, M. Krisch, D. Prabhakaran, A. T. Boothroyd, D. F. McMorrow, {\it All-In--All-Out Magnetic Order and Propagating Spin Waves in Sm$_{2}$Ir$_{2}$O$_{7}$}, Phys. Rev. Lett. {\bf 117} 037201 (2016).
	\bibitem{Sak11} M. Sakata, T. Kagayama, K. Shimizu, K. Matsuhira, S. Takagi, M. Wakeshima, Y. Hinatsu, {\it Suppression of metal-insulator transition at high pressure and pressure-induced magnetic ordering in pyrochlore oxide Nd$_{2}$Ir$_{2}$O$_{7}$}, Phys. Rev. B {\bf 83} 041102(R) (2011).	
	\bibitem{Taf12} F. F. Tafti, J. J. Ishikawa, A. McCollam, S. Nakatsuji, S. R. Julian, {\it Pressure-tuned insulator to metal transition in Eu$_{2}$Ir$_{2}$O$_{7}$}, Phys. Rev. B {\bf 85} 205104 (2012).
	\bibitem{Ued15} K. Ueda, J. Fujioka, C. Terakura, Y. Tokura, {\it Pressure and magnetic field effects on metal-insulator transitions of bulk and domain wall states in pyrochlore iridates}, Phys. Rev. B {\bf 92} 121110(R) (2015).
	\bibitem{Pra16} G. Prando, R. Dally, W. Schottenhamel, Z. Guguchia, S.-H. Baek, R. Aeschlimann, A. U. B. Wolter, S. D. Wilson, B. B\"uchner, M. J. Graf, {\it Influence of hydrostatic pressure on the bulk magnetic properties of Eu$_{2}$Ir$_{2}$O$_{7}$}, Phys. Rev. B {\bf 93} 104422 (2016).
	\bibitem{Tel19} P. Telang, K. Mishra, G. Prando, A. K. Sood, S. Singh, {\it Anomalous lattice contraction and emergent electronic phases in Bi-doped Eu$_{2}$Ir$_{2}$O$_{7}$}, Phys. Rev. B {\bf 99} 201112(R) (2019).
	\bibitem{SI} See the Supplemental Material, which includes Refs.~\cite{Pra13b,Pra15,Pra15b}, for general technical details on $\mu^{+}$SR as well as for a discussion of the data analysis for the measurements under applied pressure.
	\bibitem{Pra13b} G. Prando, P. Bonf\`a, G. Profeta, R. Khasanov, F. Bernardini, M. Mazzani, E. M. Br\"uning, A. Pal, V. P. S. Awana, H.-J. Grafe, B. B\"uchner, R. De Renzi, P. Carretta, S. Sanna, {\it Common effect of chemical and external pressures on the magnetic properties of RCoPO (R = La, Pr)}, Phys. Rev. B {\bf 87} 064401 (2013).
	\bibitem{Pra15} G. Prando, G. Profeta, A. Continenza, R. Khasanov, A. Pal, V. P. S. Awana, B. B\"uchner, S. Sanna, {\it Common effect of chemical and external pressures on the magnetic properties of RCoPO (R = La, Pr, Nd, Sm). II.}, Phys. Rev. B {\bf 92} 144414 (2015).
	\bibitem{Pra15b} G. Prando, T. Hartmann, W. Schottenhamel, Z. Guguchia, S. Sanna, F. Ahn, I. Nekrasov, C. G. F. Blum, A. U. B. Wolter, S. Wurmehl, R. Khasanov, I. Eremin, B. B\"uchner, {\it Mutual Independence of Critical Temperature and Superfluid Density under Pressure in Optimally Electron-Doped Superconducting LaFeAsO$_{1-x}$F$_{x}$}, Phys. Rev. Lett. {\bf 114} 247004 (2015).	
	\bibitem{Yam14} Y. Yamaji, M. Imada, {\it Metallic Interface Emerging at Magnetic Domain Wall of Antiferromagnetic Insulator: Fate of Extinct Weyl Electrons}, Phys. Rev. X {\bf 4}, 021035 (2014).
	\bibitem{Ma15} E. Y. Ma, Y.-T. Cui, K. Ueda, S. Tang, K. Chen, N. Tamura, P. M. Wu, J. Fujioka,	Y. Tokura, Z.-X. Shen, {\it Mobile metallic domain walls in an all-in--all-out magnetic insulator}, Science {\bf 350}, 538 (2015).
	\bibitem{Tar15} S. Tardif, S. Takeshita, H. Ohsumi, J.-I. Yamaura, D. Okuyama, Z. Hiroi, M. Takata, T.-H. Arima, {\it All-In--All-Out Magnetic Domains: X-Ray Diffraction Imaging and Magnetic Field Control}, Phys. Rev. Lett. {\bf 114} 147205 (2015).
	\bibitem{Hir17} H. T. Hirose, J.-I. Yamaura, Z. Hiroi, {\it Robust ferromagnetism carried by antiferromagnetic domain walls}, Sci. Rep. {\bf 7} 42440 (2017).
	\bibitem{Kim18} W. J. Kim, J. H. Gruenewald, T. Oh, S. Cheon, B. Kim, O. B. Korneta, H. Cho, D. Lee, Y. Kim, M. Kim, J.-G. Park, B.-J. Yang, A. Seo, T. W. Noh, {\it Unconventional anomalous Hall effect from antiferromagnetic domain walls of Nd$_{2}$Ir$_{2}$O$_{7}$ thin films}, Phys. Rev. B {\bf 98} 125103 (2018).	
	\bibitem{Blu99} S. J. Blundell, {\it Spin-polarized muons in condensed matter physics}, Contemp. Phys. {\bf 40} 175 (1999).
	\bibitem{Yao11} A. Yaouanc, P. Dalmas de R\'eotier, {\it Muon Spin Rotation, Relaxation, and Resonance: Applications to Condensed Matter}, Oxford University Press, Oxford (2011).
	\bibitem{Blu12} S. J. Blundell, A. J. Steele, T. Lancaster, J. D. Wright, F. L. Pratt, {\it A Bayesian approach to magnetic moment determination using $\mu$SR}, Phys. Procedia {\bf 30} 113 (2012).
	\bibitem{Mol13} J. S. M\"oller, P. Bonf\`a, D. Ceresoli, F. Bernardini, S. J. Blundell,	T. Lancaster, R. De Renzi, N. Marzari, I. Watanabe, S. Sulaiman, M. I. Mohamed-Ibrahim, {\it Playing quantum hide-and-seek with the muon: localizing muon stopping sites}, Phys. Scr. {\bf 88} 068510 (2013).
	\bibitem{Bon16} P. Bonf\`a, R. De Renzi, {\it Toward the Computational Prediction of Muon Sites and Interaction Parameters}, Journ. Phys. Soc. Jpn. {\bf 85} 091014 (2016).
	\bibitem{Sta71} H. E. Stanley, {\it Introduction to phase transitions and critical phenomena}, Oxford University Press (1971).
	\bibitem{Kla08} H.-H. Klauss, H. Luetkens, R. Klingeler, C. Hess, F. J. Litterst, M. Kraken, M. M. Korshunov, I. Eremin, S.-L. Drechsler, R. Khasanov, A. Amato, J. Hamann-Borrero, N. Leps, A. Kondrat, G. Behr, J. Werner, B. B\"uchner, {\it Commensurate Spin Density Wave in LaFeAsO: A Local Probe Study}, Phys. Rev. Lett. {\bf 101} 077005 (2008).
	\bibitem{Pra13} G. Prando, O. Vakaliuk, S. Sanna, G. Lamura, T. Shiroka, P. Bonf\`a, P. Carretta, R. De Renzi, H.-H. Klauss, C. G. F. Blum, S. Wurmehl, C. Hess, B. B\"uchner, {\it Role of in-plane and out-of-plane dilution in CeFeAsO: Charge doping versus disorder}, Phys. Rev. B {\bf 87} 174519 (2013).
	\bibitem{Mat15} P. Materne, S. Kamusella, R. Sarkar, T. Goltz, J. Spehling, H. Maeter, L. Harnagea, S. Wurmehl, B. B\"uchner, H. Luetkens, C. Timm, H.-H. Klauss, {\it Coexistence of superconductivity and magnetism in Ca$_{1-x}$Na$_{x}$Fe$_{2}$As$_{2}$: Universal suppression of the magnetic order parameter in 122 iron pnictides}, Phys. Rev. B {\bf 92} 134511 (2015).
	\bibitem{Fra16} B. A. Frandsen, L. Liu, S. C. Cheung, Z. Guguchia, R. Khasanov, E. Morenzoni, T. J. S. Munsie, A. M. Hallas, M. N. Wilson, Y. Cai, G. M. Luke, B. Chen, W. Li, C. Jin, C. Ding, S. Guo, F. Ning, T. U. Ito, W. Higemoto, S. J. L. Billinge, S. Sakamoto, A. Fujimori, T. Murakami, H. Kageyama, J. A. Alonso, G. Kotliar, M. Imada, Y. J. Uemura, {\it Volume-wise destruction of the antiferromagnetic Mott insulating state through quantum tuning}, Nat. Commun. {\bf 7} 12519 (2016).
	\bibitem{Wil16} M. N. Wilson, T. J. Williams, Y.-P. Cai, A. M. Hallas, T. Medina, T. J. Munsie, S. C. Cheung, B. A. Frandsen, L. Liu, Y. J. Uemura, G. M. Luke, {\it Antiferromagnetism and hidden order in isoelectronic doping of URu$_{2}$Si$_{2}$}, Phys. Rev. B {\bf 93} 064402 (2016).
	\bibitem{Cha95} P. M. Chaikin, T. C. Lubensky, {\it Principles of condensed matter physics}, Cambridge University Press (1995).
	\bibitem{Abr74} F. F. Abraham, {\it Homogeneous nucleation theory}, Academic Press (1974).
	\bibitem{Blu06} S. J. Blundell, K. M. Blundell, {\it Concepts in thermal physics}, Oxford University Press (2006).
	\bibitem{Bra12} H.-B. Braun, {\it Topological effects in nanomagnetism: from superparamagnetism to chiral quantum solitons}, Adv. Phys. {\bf 61} 1 (2012).
	\bibitem{Zan18} J. Zang, V. Cros, A. Hoffmann (eds.), {\it Topology in magnetism}, Springer (2018).
\end{thebibliography}
\end{document}